\documentclass[pra,showpacs]{revtex4}
\usepackage{graphicx}
\usepackage{dcolumn}
\usepackage{bm}

\begin{document}

\title{Dynamics of Bloch vector in thermal Jaynes-Cummings model}

\date{\today}

\author{Hiroo Azuma}
\altaffiliation{On leave from
Research Center for Quantum Information Science,
Tamagawa University Research Institute,
6-1-1 Tamagawa-Gakuen, Machida-shi, Tokyo 194-8610, Japan}
\email[Electronic address: ]{hiroo.azuma@m3.dion.ne.jp}
\affiliation{2-1-12 MinamiFukunishi-cho,
Oe, NishiKyo-ku, Kyoto-shi, Kyoto 610-1113, Japan}

\begin{abstract}
In this paper, we investigate the dynamics of the Bloch vector of
a single two-level atom which interacts with a single quantized electromagnetic field mode
according to the Jaynes-Cummings model,
where the field is initially prepared in a thermal state.
The time evolution of the Bloch vector $\mbox{\boldmath $S$}(t)$ seems to be in complete disorder
because of the thermal distribution of the initial state of the field.
Both the norm and the direction of $\mbox{\boldmath $S$}(t)$ oscillate hard
and their periods seem infinite.
We observe that the trajectory of the time evolution of $\mbox{\boldmath $S$}(t)$
in the two- or three-dimensional space
does not form a closed path.
To remove the fast frequency oscillation from the trajectory,
we take the time-average of the Bloch vector $\mbox{\boldmath $S$}(t)$.
We examine the histogram of $\{S_{z}(n\Delta t)|n=0,1,...,N\}$
for small $\Delta t$ and large $N$.
It represents an absolute value of a derivative of the inverse function of $S_{z}(t)$.
(When the inverse function of $y=S_{z}(t)$ is a multi-valued function,
the histogram represents a summation of the absolute values of its derivatives
at points whose real parts are equal to $y$ on the Riemann surface.)
We examine the dependence of the variance of the histogram on the temperature
of the field.
We estimate the lower bound of the entanglement between the atom and the field.
\end{abstract}


\maketitle

\section{\label{section-introduction}Introduction}
The Jaynes-Cummings model (JCM) is a solvable quantum mechanical model of a single
two-level atom in a single electromagnetic field mode
\cite{Jaynes,Shore,Walls,Louisell,Schleich}.
This model is originally designed for studying a spontaneous emission.
The interaction term is obtained by the rotating wave approximation.
In this interaction, each photon creation causes an atomic de-excitation
and each photon annihilation causes an atomic excitation.

If the photon number is sharply defined in the initial state of the field,
the JCM shows the Rabi oscillations in the populations of the atomic levels.
If the initial state of the field is a coherent state,
the oscillation of the mean photon number collapses and revives in the JCM.
In this way, the JCM reveals the quantum natures of the radiation.

Because the JCM is exactly solvable,
it is investigated from various viewpoints.
The JCM whose boson field is prepared in a thermal state is discussed in
Refs.~\cite{Arroyo-Correa,Chumakov,Klimov}.
Thermodynamics of the JCM is discussed in Ref.~\cite{Liu}.
In this analysis, the grand partition function of both the atom and the boson field is considered.
Extended JCMs are studied as dissipative models
\cite{Murao,Uchiyama}.

Recently, the JCM has been used for describing the evolution of the entanglement
between the atom and the field
\cite{Bose,Scheel}.
In Refs.~\cite{Bose,Scheel}, the electromagnetic field is assumed to be
initially prepared in a thermal state.
In these papers, the JCM is regarded as a source of the entanglement between the atom and the field.
In Ref.~\cite{Kim}, this idea is advanced and generation of the entanglement between two atoms
interacting with a single-mode thermal field according to the JCM is discussed.
In this model, the atoms which are initially in a separable state obtain the entanglement
through the time evolution.

In Ref.~\cite{Rendell}, evolution of the entanglement between an atom and a single-mode field
described with the JCM under phase damping is studied.
In Ref.~\cite{Larson}, an atom interacting with two cavity modes is considered
and it is shown that this system can be reduced to the JCM.
Using this system, generation of entangled coherent states is discussed.
In Refs.~\cite{El-Orany,Bradler}, the entanglement in an extended JCM where two atoms interacts
with a single-mode field is studied.

As an interesting phenomenon related with the time evolution of the entanglement in the JCM,
so called sudden death effect is studied \cite{Yu-Eberly1,Yonac,Yu-Eberly2}.
In these references, two isolated atoms, each of which is located in its own Jaynes-Cummings
cavity, are assumed.
If these atoms are in a certain entangled state initially,
the entanglement disappears in a finite time.
This phenomenon is experimentally demonstrated in Ref.~\cite{Almeida}.
In Ref.~\cite{Sainz}, an attempt to find invariant entanglement among atoms and fields
in two isolated JCM is done.

In this paper, we consider the dynamics of the Bloch vector of a two-level atom
interacting with a single mode boson according to the JCM,
where the initial state of the boson is in thermal equilibrium.
The time evolution of the Bloch vector $\mbox{\boldmath $S$}(t)$
seems to be in a state of complete disorder because of the thermal distribution
of the initial state of the boson field.
The trajectory of $\mbox{\boldmath $S$}(t)$ oscillates hard and seems to wander without a purpose.
We try to find a property that characterizes this confused movement of $\mbox{\boldmath $S$}(t)$.

If $S_{x}(0)\neq 0$ or $S_{y}(0)\neq 0$, the time evolution of the Bloch vector $\mbox{\boldmath $S$}(t)$
draws a trajectory in the two- or three-dimensional space.
We observe that this trajectory does not form a closed path.
To remove the fast frequency oscillation from the trajectory,
we take the time-average of the $\mbox{\boldmath $S$}(t)$ as
$\langle \mbox{\boldmath $S$} \rangle=\lim_{T\to\infty}(1/T)\int_{0}^{T}dt\:\mbox{\boldmath $S$}(t)$.
We find that $\langle S_{x}\rangle$ and $\langle S_{y}\rangle$ are equal to zero
$\forall \beta$, $\omega$ and $\omega_{0}$, where $\beta$ is an inverse of the temperature
for the initial thermal state of the field,
$\omega$ is an angular frequency of the field and $\hbar\omega_{0}$ is an energy gap of the
two-level atom.
$\langle S_{z}\rangle$ takes a certain function of $\beta$, $\omega$ and $\omega_{0}$.

We take the histogram of samples $\{S_{z}(n\Delta t)|n=0,1,...,N\}$
for small $\Delta t$ and large $N$.
We understand that it represents an absolute value of a derivative of the inverse function of $S_{z}(t)$.
(When the inverse function of $y=S_{z}(t)$ is a multi-valued function,
the histogram represents a summation of the absolute values of its derivatives
$|(d/dy)S_{z}^{-1}|$
at points whose real parts are equal to $y$ on the Riemann surface,
where $S_{z}^{-1}$ is the inverse function of $y=S_{z}(t)$.)
We approximate obtained histograms at a high temperature by the probability function of the normal
distribution.
We examine the dependence of the variance of the samples on the temperature $\beta$.

We examine the time evolution of the entanglement between the atom and the field in our model
by estimating the entanglement of formation of the density matrix which lies on the $2\times 2$ dimensional
projected subspace of the atom and the field.
Because this projection is an local operation,
the entanglement of formation computed in the reduced $2\times 2$ dimensional subspace gives
the lower bound of the entanglement of formation of the whole system
(the atom and the field).

This paper is organized as follows:
In Sec.~\ref{section-evolution-Bloch-vector}, we derive an equation
which governs the time evolution of the Bloch vector $\mbox{\boldmath $S$}(t)$.
In Sec.~\ref{section-trajectory-Bloch-vector}, we examine the trajectory of $\mbox{\boldmath $S$}(t)$
and observe that it does not form a closed path.
In Sec.~\ref{section-time-average-Bloch-vector-Delta-omega-eq-zero},
we take the time-average of $\mbox{\boldmath $S$}(t)$ for the case where the field is
resonant with the atom.
In Sec.~\ref{section-histogram-data-Szt-sampled-intervals-Deltat},
we take the histogram of data of $S_{z}(t)$ sampled at intervals of $\Delta t$.
In Sec.~\ref{section-time-average-Bloch-vector-Delta-omega-neq-zero},
we take the time-average of $\mbox{\boldmath $S$}(t)$ for the non-resonant case.
In Sec.~\ref{section-entanglement}, we consider the lower bound of the entanglement of formation
between the atom and the field.
In Sec.~\ref{section-Discussions}, we give brief discussions.

\section{\label{section-evolution-Bloch-vector}The time evolution of the Bloch vector}
In this section, we give the equation of the time evolution of the Bloch vector of the atom.
The Jaynes-Cummings model is a system that is described by the following Hamiltonian:
\begin{equation}
H=\frac{\hbar}{2}\omega_{0}\sigma_{z}+\hbar\omega a^{\dagger}a
+\hbar g(\sigma_{+}a+\sigma_{-}a^{\dagger}),
\label{JCM-Hamiltonian}
\end{equation}
where
\begin{equation}
\sigma_{\pm}=\frac{1}{2}(\sigma_{x}\pm i\sigma_{y}),
\label{definition-sigma+-}
\end{equation}
\begin{equation}
\sigma_{x}=
\left(
\begin{array}{cc}
0 & 1 \\
1 & 0
\end{array}
\right),
\quad
\sigma_{y}=
\left(
\begin{array}{cc}
0 & -i \\
i & 0
\end{array}
\right),
\quad
\sigma_{z}=
\left(
\begin{array}{cc}
1 & 0 \\
0 & -1
\end{array}
\right),
\end{equation}
and
$[a,a^{\dagger}]=1$.
The Pauli matrices ($\sigma_{i}$, $i=x,y,z$) are operators of the atom
and $a$ and $a^{\dagger}$
are operators of the field.
In this paper, we assume that $g$ is a constant,
so that it does not depend on $\omega_{0}$ or $\omega$.
Let us divide $H$ as follows \cite{Louisell,Schleich}:
\begin{eqnarray}
H&=&\hbar(C_{1}+C_{2}), \label{operator-H} \\
C_{1}&=&\omega(\frac{1}{2}\sigma_{z}+a^{\dagger}a), \label{operator-C1} \\
C_{2}&=&g(\sigma_{+}a+\sigma_{-}a^{\dagger})-\frac{\Delta\omega}{2}\sigma_{z}, \label{operator-C2}
\end{eqnarray}
where $\Delta\omega=\omega-\omega_{0}$.
We can confirm
\begin{equation}
[C_{1},C_{2}]=0.
\label{commutation-relation-C1-C2}
\end{equation}

Because $C_{1}$ can be diagonalized at ease,
we take the following interaction picture.
We write a state vector of the whole system in the Schr{\"o}dinger picture as $|\psi(t)\rangle$.
A state vector in the interaction picture is defined by
\begin{equation}
|\psi_{\mbox{\scriptsize I}}(t)\rangle=\exp(iC_{1}t)|\psi(t)\rangle.
\label{definition-interaction-rep}
\end{equation}
(We assume $|\psi_{\mbox{\scriptsize I}}(0)\rangle=|\psi(0)\rangle$.)
Because of Eq.~(\ref{commutation-relation-C1-C2}),
the time evolution of $|\psi_{\mbox{\scriptsize I}}(t)\rangle$ is given by
\begin{equation}
|\psi_{\mbox{\scriptsize I}}(t)\rangle=U(t)|\psi(0)\rangle,
\label{evolution-interaction-rep}
\end{equation}
where
\begin{equation}
U(t)=\exp(-iC_{2}t).
\label{evolution-unitary-operator}
\end{equation}

We define the density operator of the initial state of the atom and the boson field as
\begin{equation}
\rho_{\mbox{\scriptsize AF}}(0)
=\rho_{\mbox{\scriptsize A}}(0)\otimes\rho_{\mbox{\scriptsize F}},
\label{whole-system-initial-state}
\end{equation}
\begin{equation}
\rho_{\mbox{\scriptsize A}}(0)
=\sum_{i,j\in\{0,1\}}\rho_{\mbox{\scriptsize A},ij}(0)
|i\rangle_{\mbox{\scriptsize A}}{}_{\mbox{\scriptsize A}}\langle j|,
\label{atom-initial-state}
\end{equation}
\begin{eqnarray}
\rho_{\mbox{\scriptsize F}}
&=&
\frac{\exp(-\beta\hbar\omega a^{\dagger}a)}{\mbox{Tr}\exp(-\beta\hbar\omega a^{\dagger}a)} \nonumber \\
&=&
(1-e^{-\beta\hbar\omega})\exp(-\beta\hbar\omega a^{\dagger}a),
\label{field-initial-state}
\end{eqnarray}
where
\begin{equation}
|0\rangle_{\mbox{\scriptsize A}}
=
\left(
\begin{array}{c}
1 \\
0
\end{array}
\right),
\quad
|1\rangle_{\mbox{\scriptsize A}}
=
\left(
\begin{array}{c}
0 \\
1
\end{array}
\right),
\label{atom-basis-vector}
\end{equation}
and $\beta$ is an inverse of the temperature.
(The indices $\mbox{A}$ and $\mbox{F}$ imply the atom and the field, respectively.)

The density operator of the atom in the interaction picture evolves according to
\begin{equation}
\rho_{\mbox{\scriptsize A}}(t)
=\sum_{i,j\in\{0,1\}}\rho_{\mbox{\scriptsize A},ij}(t)
|i\rangle_{\mbox{\scriptsize A}}{}_{\mbox{\scriptsize A}}\langle j|,
\label{atom-evolution1}
\end{equation}
\begin{eqnarray}
\rho_{\mbox{\scriptsize A},ij}(t)
&=&\sum_{k,l\in\{0,1\}}\rho_{\mbox{\scriptsize A},kl}(0)
A_{kl,ij}(t) \nonumber \\
&&\quad\quad
\mbox{for $i,j\in\{0,1\}$},
\label{atom-evolution2}
\end{eqnarray}
\begin{equation}
A_{kl,ij}(t)
={}_{\mbox{\scriptsize A}}\langle i|\mbox{Tr}_{\mbox{\scriptsize F}}
[U(t)(|k\rangle_{\mbox{\scriptsize A}}{}_{\mbox{\scriptsize A}}\langle l|\otimes \rho_{\mbox{\scriptsize F}})
U^{\dagger}(t)]|j\rangle_{\mbox{\scriptsize A}},
\label{atom-evolution3}
\end{equation}
where $\mbox{Tr}_{\mbox{\scriptsize F}}$ means a partial trace over the field.
Because $\rho_{\mbox{\scriptsize A},10}(t)=\rho_{\mbox{\scriptsize A},01}(t)^{*}$ and
$\rho_{\mbox{\scriptsize A},11}(t)=1-\rho_{\mbox{\scriptsize A},00}(t)$,
we examine $\rho_{\mbox{\scriptsize A},00}(t)$ and
$\rho_{\mbox{\scriptsize A},01}(t)$ only.
We can derive $A_{kl,ij}(t)$ as follows:
The unitary evolution operator of the whole system $U(t)$ given by
Eqs.~(\ref{operator-C2}) and (\ref{evolution-unitary-operator})
is rewritten as
\begin{eqnarray}
U(t)
&=&
\exp[-it
\left(
\begin{array}{cc}
-\Delta\omega/2 & ga \\
ga^{\dagger} & \Delta\omega/2
\end{array}
\right)
] \nonumber \\
&=&
\sum_{n=0}^{\infty}\frac{(-1)^{n}t^{2n}}{(2n)!}
\left(
\begin{array}{cc}
(D+g^{2})^{n} & 0 \\
0 & D^{n}
\end{array}
\right) \nonumber \\
&&\quad\quad
+
\sum_{n=0}^{\infty}\frac{i(-1)^{n}t^{2n+1}}{(2n+1)!}
\left(
\begin{array}{cc}
-\Delta\omega(D+g^{2})^{n}/2 & gaD^{n} \\
ga^{\dagger}(D+g^{2})^{n} & \Delta\omega D^{n}/2
\end{array}
\right) \nonumber \\
&=&
\left(
\begin{array}{cc}
u_{00} & u_{01} \\
u_{10} & u_{11}
\end{array}
\right), \label{explicit-time-evolution-op}
\end{eqnarray}
where
\begin{equation}
D=(\frac{\Delta\omega}{2})^{2}+g^{2}a^{\dagger}a,
\label{definition-op-D}
\end{equation}
\begin{eqnarray}
u_{00}&=&
\cos(t\sqrt{D+g^{2}})
-\frac{i}{2}\Delta\omega\frac{\sin(t\sqrt{D+g^{2}})}{\sqrt{D+g^{2}}}, \nonumber \\
u_{01}&=&
iga\frac{\sin(t\sqrt{D})}{\sqrt{D}}, \nonumber \\
u_{10}&=&
iga^{\dagger}\frac{\sin(t\sqrt{D+g^{2}})}{\sqrt{D+g^{2}}}, \nonumber \\
u_{11}&=&
\cos(t\sqrt{D})
+\frac{i}{2}\Delta\omega\frac{\sin(t\sqrt{D})}{\sqrt{D}}.
\label{elements-time-evolution-op}
\end{eqnarray}
From Eqs.~(\ref{atom-evolution3}), (\ref{explicit-time-evolution-op}),
(\ref{definition-op-D}) and (\ref{elements-time-evolution-op}), we obtain
\begin{eqnarray}
A_{00,00}(t)&=&
(1-e^{-\beta\hbar\omega})
\sum_{n=0}^{\infty}
\frac{(\Delta\omega/2)^{2}+g^{2}(n+1)\cos^{2}(\sqrt{\tilde{D}(n+1)}t)}{\tilde{D}(n+1)}
e^{-n\beta\hbar\omega}, \nonumber \\
A_{11,00}(t)&=&
(1-e^{-\beta\hbar\omega})
\sum_{n=1}^{\infty}
g^{2}n
\frac{\sin^{2}(\sqrt{\tilde{D}(n)}t)}{\tilde{D}(n)}
e^{-n\beta\hbar\omega}, \nonumber \\
A_{01,01}(t)&=&
(1-e^{-\beta\hbar\omega})
\sum_{n=0}^{\infty}
[\cos(\sqrt{\tilde{D}(n+1)}t)
-\frac{i}{2}\Delta\omega\frac{\sin(\sqrt{\tilde{D}(n+1)}t)}{\sqrt{\tilde{D}(n+1)}}] \nonumber \\
&&\quad\quad\times
[\cos(\sqrt{\tilde{D}(n)}t)
-\frac{i}{2}\Delta\omega\frac{\sin(\sqrt{\tilde{D}(n)}t)}{\sqrt{\tilde{D}(n)}}]e^{-n\beta\hbar\omega},
\label{A-elements-explicit-form1}
\end{eqnarray}
\begin{equation}
A_{01,00}(t)=A_{10,00}(t)=A_{00,01}(t)=A_{10,01}(t)=A_{11,01}(t)=0,
\label{A-elements-explicit-form2}
\end{equation}
where
\begin{equation}
\tilde{D}(n)=(\frac{\Delta\omega}{2})^{2}+g^{2}n.
\label{definition-tilde-D}
\end{equation}
(The trace over the field is taken by the basis vectors of the photon
number states.)

We introduce the Bloch vector $\mbox{\boldmath $S$}(t)=(S_{x}(t),S_{y}(t),S_{z}(t))$
which is given by
\begin{equation}
\rho_{\mbox{\scriptsize A}}(t)
=
\frac{1}{2}(\mbox{\boldmath $I$}+\mbox{\boldmath $S$}(t)\cdot\mbox{\boldmath $\sigma$}),
\label{definition-Bloch-vector}
\end{equation}
where $\mbox{\boldmath $I$}$ is the identity operator and
$\mbox{\boldmath $\sigma$}=(\sigma_{x},\sigma_{y},\sigma_{z})$.
Because $\rho_{\mbox{\scriptsize A}}(t)^{\dagger}=\rho_{\mbox{\scriptsize A}}(t)$ and
$\rho_{\mbox{\scriptsize A}}(t)\geq 0$,
$\mbox{\boldmath $S$}(t)$ is a real vector and satisfies $|\mbox{\boldmath $S$}(t)|^{2}\leq 1$.
From Eqs.~(\ref{atom-evolution2}), (\ref{A-elements-explicit-form2}) and (\ref{definition-Bloch-vector}),
we obtain
\begin{eqnarray}
\mbox{\boldmath $S$}(t)
&=&
{\cal L}_{\Delta\omega}(t)\mbox{\boldmath $S$}(0) \nonumber \\
&=&
\left(
\begin{array}{ccc}
L_{\Delta\omega}^{(1)}(t) & L_{\Delta\omega}^{(2)}(t) & 0 \\
-L_{\Delta\omega}^{(2)}(t) & L_{\Delta\omega}^{(1)}(t) & 0 \\
0 & 0 & L_{\Delta\omega}^{(3)}(t)
\end{array}
\right)
\mbox{\boldmath $S$}(0)
+
\left(
\begin{array}{c}
0 \\
0 \\
L_{\Delta\omega}^{(4)}(t)
\end{array}
\right),
\label{equation-time-evolution-Bloch-vector-general}
\end{eqnarray}
where
\begin{eqnarray}
L_{\Delta\omega}^{(1)}(t)&=&\mbox{Re}[A_{01,01}(t)], \nonumber \\
L_{\Delta\omega}^{(2)}(t)&=&\mbox{Im}[A_{01,01}(t)], \nonumber \\
L_{\Delta\omega}^{(3)}(t)&=&A_{00,00}(t)-A_{11,00}(t), \nonumber \\
L_{\Delta\omega}^{(4)}(t)&=&A_{00,00}(t)+A_{11,00}(t)-1.
\label{elements-equation-evolution-Bloch-vector-general}
\end{eqnarray}
This is the equation of the time evolution of \mbox{\boldmath $S$}(t).

\section{\label{section-trajectory-Bloch-vector}The trajectory of the Bloch vector}
In this section, we observe the trajectory of the time evolution of $\mbox{\boldmath $S$}(t)$.
To simplify the discussion, we concentrate on the case
where the field is resonant with the atom, that is,
$\Delta\omega=0$.
Furthermore, we regard $\omega_{0}$ as a constant.
Thus, the model has two variables, $t$ and $\beta$.
We replace $|g|t$ with $t$ and $\beta\hbar\omega(=\beta\hbar\omega_{0})$ with $\beta$,
where we assume $g\neq 0$ and $\omega_{0}\neq 0$.
They imply that the time $t$ is in units of $|g|^{-1}$
and the inverse of the temperature $\beta$ is in units of $(\hbar\omega_{0})^{-1}$.
From the above assumptions and
Eqs.~(\ref{A-elements-explicit-form1}), (\ref{definition-tilde-D}),
(\ref{equation-time-evolution-Bloch-vector-general}) and
(\ref{elements-equation-evolution-Bloch-vector-general}),
we obtain the following equation:
\begin{eqnarray}
\mbox{\boldmath $S$}(t)
&=&
{\cal L}(t)\mbox{\boldmath $S$}(0) \nonumber \\
&=&
\left(
\begin{array}{ccc}
L_{1}(t) & 0 & 0 \\
0 & L_{1}(t) & 0 \\
0 & 0 & L_{3}(t)
\end{array}
\right)
\mbox{\boldmath $S$}(0)
+
\left(
\begin{array}{c}
0\\
0\\
L_{4}(t)
\end{array}
\right),
\label{evolution-Bloch-vector-resonant}
\end{eqnarray}
where
\begin{eqnarray}
L_{1}(t)&=&
(1-e^{-\beta})
\sum_{n=0}^{\infty}
\cos(\sqrt{n+1}t)\cos(\sqrt{n}t)e^{-n\beta}, \label{explicit-form-L1} \\
L_{3}(t)&=&
\frac{1}{2}(1-e^{-\beta})
+\frac{e^{2\beta}-1}{2e^{\beta}}
\sum_{n=1}^{\infty}
\cos(2\sqrt{n}t)e^{-n\beta}, \label{explicit-form-L3} \\
L_{4}(t)&=&
-\frac{1}{2}(1-e^{-\beta})
+\frac{(e^{\beta}-1)^{2}}{2e^{\beta}}
\sum_{n=1}^{\infty}
\cos(2\sqrt{n}t)e^{-n\beta}. \label{explicit-form-L4}
\end{eqnarray}
(We note $\mbox{Im}[A_{01,01}(t)]=0$ for $\Delta\omega=0$.
In the algebraic form of $\mbox{Im}[A_{01,01}(t)]$,
summations of
\\
$\cos(\sqrt{n+1}|g|t)\sin(\sqrt{n}|g|t)e^{-n\beta\hbar\omega}/\sqrt{n}$
and
$\cos(\sqrt{n}|g|t)\sin(\sqrt{n+1}|g|t)e^{-n\beta\hbar\omega}/\sqrt{n+1}$
for the index $n$ from $0$ to $\infty$ appear,
where $\left.\sin(\sqrt{n}|g|t)/\sqrt{n}\right|_{n=0}=|g|t$.
However, they converge to finite values.
Similar discussions are explained in Sec.~\ref{section-time-average-Bloch-vector-Delta-omega-eq-zero}.)
Looking at Eq.~(\ref{evolution-Bloch-vector-resonant}),
we understand that the trajectory (a curve in the three-dimensional space of $\mbox{\boldmath $S$}$
parameterized by $t$) is uniquely determined by the initial state $\mbox{\boldmath $S$}(0)$
for given $\beta$.

\begin{figure}
\includegraphics[scale=1.0]{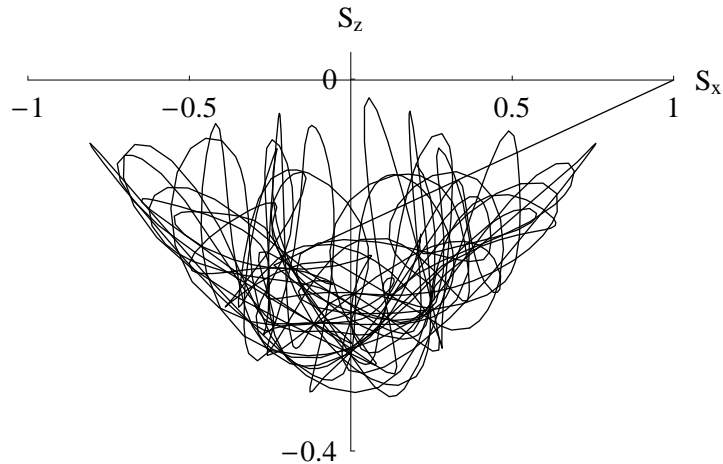}
\caption{The trajectory of $\mbox{\boldmath $S$}(t)$ ($0\leq t\leq 100$)
whose initial state and temperature are given by $\mbox{\boldmath $S$}(0)=(1,0,0)$
and $\beta=0.5$, respectively.
(We assume $\Delta\omega=0$.)
The horizontal and vertical lines represent $S_{x}$ and $S_{z}$, respectively.
They are dimensionless quantities.
In this case, the trajectory lies on the $xz$-plane.
In the numerical calculation of $L_{1}(t)$ and $L_{4}(t)$ defined in
Eqs.~(\ref{explicit-form-L1}) and (\ref{explicit-form-L4}),
the summations of the index $n$ is carried out up to $500$.}
\label{Figure01}
\end{figure}

Figure~\ref{Figure01} shows a trajectory of $\mbox{\boldmath $S$}(t)$
whose initial state and temperature are given by $\mbox{\boldmath $S$}(0)=(1,0,0)$
and $\beta=0.5$, respectively.
In this case, $S_{y}(t)=0$ for $t\geq 0$ and the curve of $\mbox{\boldmath $S$}(t)$
lies on the $xz$-plane.
The trajectory of Fig.~\ref{Figure01} oscillates hard and seems to be in complete disorder.
This is because $\mbox{\boldmath $S$}(t)$ is a tuple of superpositions of
$\cos(\sqrt{n+1}t)\cos(\sqrt{n}t)e^{-n\beta}$ and
$\cos(2\sqrt{n}t)e^{-n\beta}$
for $n=0,1,2,...$ as shown in Eqs.~(\ref{explicit-form-L1}) and (\ref{explicit-form-L4}).

\begin{figure}
\includegraphics[scale=1.8]{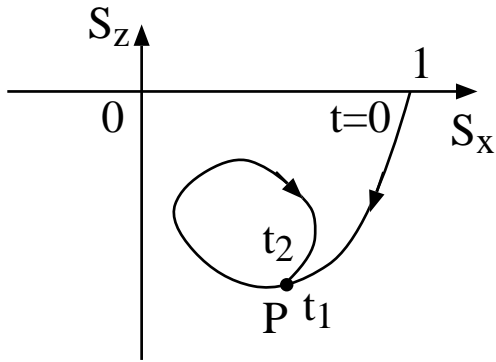}
\caption{The trajectory of $\mbox{\boldmath $S$}(t)$ whose initial state is given by
$\mbox{\boldmath $S$}(0)=(1,0,0)$.
(We assume $\Delta\omega=0$.)
The horizontal and vertical lines represent $S_{x}$ and $S_{z}$, respectively.
They are dimensionless quantities.
If ${\cal L}(t_{2})={\cal L}(t_{2}-t_{1}){\cal L}(t_{1})$ $\forall t_{1}, t_{2}$,
the trajectory forms a closed loop.
If the trajectory passes a point $P$ at $t_{1}$ and reaches $P$ at $t_{2}(>t_{1})$ again,
$\mbox{\boldmath $S$}(t)$ moves along the closed loop for $t\geq t_{2}$.
This is because the evolution of $\mbox{\boldmath $S$}(t)$ is determined by
${\cal L}(t-t_{2})\mbox{\boldmath $S$}(t_{2})$.
Actually, ${\cal L}(t_{2})={\cal L}(t_{2}-t_{1}){\cal L}(t_{1})$ does not hold in general
and the trajectory does not form the closed path.}
\label{Figure02}
\end{figure}

\begin{figure}
\includegraphics[scale=1.0]{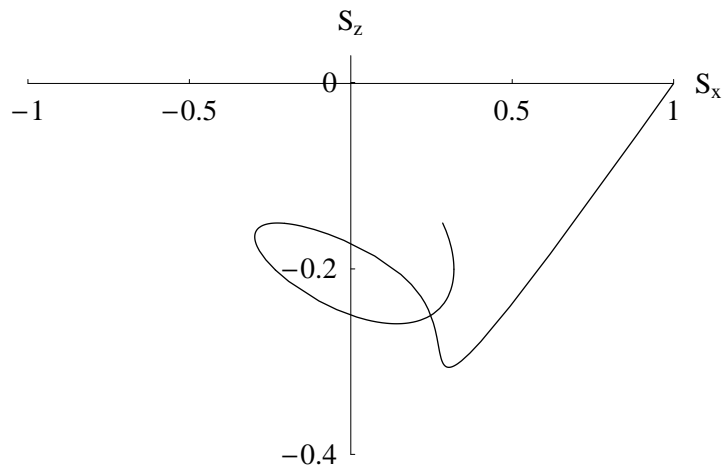}
\caption{The trajectory of $\mbox{\boldmath $S$}(t)$ ($0\leq t\leq 6$) whose initial state
and temperature are given by $\mbox{\boldmath $S$}(0)=(1,0,0)$ and $\beta=0.5$, respectively.
(We assume $\Delta\omega=0$.)
The horizontal and vertical lines represent $S_{x}$ and $S_{z}$, respectively.
They are dimensionless quantities.
We can observe that the trajectory intersects itself and does not form a closed loop.
In the numerical calculation of $L_{1}(t)$ and $L_{4}(t)$ defined in Eqs.~(\ref{explicit-form-L1})
and (\ref{explicit-form-L4}),
the summations of the index $n$ is carried out up to 500.}
\label{Figure03}
\end{figure}

If the operator ${\cal L}(t)$ defined in Eq.~(\ref{evolution-Bloch-vector-resonant})
satisfies the relation ${\cal L}(t_{2})={\cal L}(t_{2}-t_{1}){\cal L}(t_{1})$ $\forall t_{1}, t_{2}$,
the following thing can happen.
If the trajectory reaches a point (at $t_{2}$) where it has passed before
[at $t_{1}(<t_{2})$],
it forms a closed path and $\mbox{\boldmath $S$}(t)$ moves along this closed path for $t\geq t_{2}$.
(See Fig.~\ref{Figure02}.)
However, this thing does not happen because
${\cal L}(t_{2})={\cal L}(t_{2}-t_{1}){\cal L}(t_{1})$ does not hold in general.
In fact, we can observe an example where the trajectory intersects itself
and does not form a closed path in Fig.~\ref{Figure03}.

If $S_{x}(0)\neq 0$ or $S_{y}(0)\neq 0$,
$\mbox{\boldmath $S$}(t)$ draws a trajectory in the two- or three-dimensional space.
From the above considerations, we understand that this trajectory does not form a closed loop
in general.

\section{\label{section-time-average-Bloch-vector-Delta-omega-eq-zero}
The time-average of the Bloch vector for $\Delta\omega=0$}
In this section, we take the time-average of the Bloch vector
for the case where the field is resonant with the atom ($\Delta\omega=0$).
As shown in Fig.~\ref{Figure01}, the evolution of $\mbox{\boldmath $S$}(t)$ is in complete disorder.
To remove the fast frequency oscillation,
we consider the time-average of $\mbox{\boldmath $S$}(t)$ as
\begin{equation}
\langle\mbox{\boldmath $S$}\rangle
=
\lim_{T\to\infty}\frac{1}{T}\int_{0}^{T}dt\:\mbox{\boldmath $S$}(t).
\label{time-average-Bloch-vector}
\end{equation}
From Eqs.~(\ref{evolution-Bloch-vector-resonant}), (\ref{explicit-form-L1}),
(\ref{explicit-form-L3}) and (\ref{explicit-form-L4}),
we obtain
\begin{equation}
\langle\mbox{\boldmath $S$}\rangle
=
\left(
\begin{array}{ccc}
\langle L_{1}\rangle & 0 & 0 \\
0 & \langle L_{1}\rangle & 0 \\
0 & 0 & \langle L_{3}\rangle
\end{array}
\right)
\mbox{\boldmath $S$}(0)
+
\left(
\begin{array}{c}
0 \\
0 \\
\langle L_{4}\rangle
\end{array}
\right),
\end{equation}
\begin{eqnarray}
\langle L_{1}\rangle
&=&
(1-e^{-\beta})\lim_{T\to\infty}\frac{1}{T}
\sum_{n=0}^{\infty}
[\sqrt{n+1}\sin(\sqrt{n+1}T)\cos(\sqrt{n}T) \nonumber \\
&&\quad\quad
-\sqrt{n}\cos(\sqrt{n+1}T)\sin(\sqrt{n}T)]e^{-n\beta}, \label{time-average-L1} \\
\langle L_{3}\rangle
&=&
\frac{1}{2}(1-e^{-\beta})
+\frac{e^{2\beta}-1}{4e^{\beta}}
\lim_{T\to\infty}\frac{1}{T}\sum_{n=1}^{\infty}
\frac{\sin(2\sqrt{n}T)}{\sqrt{n}}
e^{-n\beta}, \label{time-average-L3} \\
\langle L_{4}\rangle
&=&
-\frac{1}{2}(1-e^{-\beta})
+\frac{(e^{\beta}-1)^{2}}{4e^{\beta}}
\lim_{T\to\infty}\frac{1}{T}\sum_{n=1}^{\infty}
\frac{\sin(2\sqrt{n}T)}{\sqrt{n}}
e^{-n\beta}. \label{time-average-L4}
\end{eqnarray}

Let us take the limit of $T\to\infty$ in Eqs.~(\ref{time-average-L1}), (\ref{time-average-L3})
and (\ref{time-average-L4}).
First, we evaluate $\langle L_{1}\rangle$.
We can obtain the following relation about $\langle L_{1}\rangle$
from Eq.~(\ref{time-average-L1}):
\begin{equation}
-V_{1}\leq\langle L_{1}\rangle\leq V_{1},
\end{equation}
where
\begin{eqnarray}
V_{1}&=&
(1-e^{-\beta})\lim_{T\to\infty}\frac{1}{T}\sum_{n=0}^{\infty}
(\sqrt{n+1}+\sqrt{n})e^{-n\beta} \nonumber \\
&=&
\frac{e^{2\beta}-1}{e^{\beta}}
\lim_{T\to\infty}\frac{1}{T}
\sum_{n=0}^{\infty}\sqrt{n}e^{-n\beta}(\geq 0).
\label{inequality-L1}
\end{eqnarray}
From Eq.~(\ref{inequality-L1}), we can derive the following relation:
\begin{equation}
V_{1}<\frac{e^{2\beta}-1}{e^{\beta}}
\lim_{T\to\infty}\frac{1}{T}
\sum_{n=0}^{\infty}n e^{-n\beta}.
\end{equation}
Because
\begin{equation}
\sum_{n=0}^{\infty}n e^{-n\beta}
=
\frac{e^{\beta}}{(e^{\beta}-1)^{2}},
\end{equation}
we obtain
\begin{equation}
V_{1}<\lim_{T\to\infty}\frac{1}{T}\frac{e^{\beta}+1}{e^{\beta}-1}.
\end{equation}
Thus, if $\beta\neq 0$,
$V_{1}=0$ and $\langle L_{1}\rangle=0$.
Moreover, we can derive $\langle L_{1}\rangle=0$ for $\beta=0$
from Eq.~(\ref{time-average-L1}) in direct.
Hence, we obtain $\langle L_{1}\rangle=0$ $\forall\beta$.

Second, we evaluate $\langle L_{3}\rangle$.
We can obtain the following relation about $\langle L_{3}\rangle$
from Eq.~(\ref{time-average-L3}):
\begin{equation}
-V_{3}\leq
\frac{e^{2\beta}-1}{4e^{\beta}}
\lim_{T\to\infty}\frac{1}{T}\sum_{n=1}^{\infty}
\frac{\sin(2\sqrt{n}T)}{\sqrt{n}}
e^{-n\beta}
\leq V_{3},
\end{equation}
where
\begin{equation}
V_{3}=
\frac{e^{2\beta}-1}{4e^{\beta}}
\lim_{T\to\infty}\frac{1}{T}\sum_{n=1}^{\infty}
\frac{e^{-n\beta}}{\sqrt{n}}
(\geq 0).
\end{equation}
Moreover, we can derive the following relation:
\begin{equation}
V_{3}<
\frac{e^{2\beta}-1}{4e^{\beta}}
\lim_{T\to\infty}\frac{1}{T}\sum_{n=1}^{\infty}
e^{-n\beta}.
\end{equation}
Because $\sum_{n=1}^{\infty}e^{-n\beta}=1/(e^{\beta}-1)$,
we obtain
\begin{equation}
V_{3}<\lim_{T\to\infty}\frac{1}{T}\frac{e^{\beta}+1}{4e^{\beta}}=0.
\end{equation}
Thus, we obtain
\begin{equation}
\langle L_{3}\rangle=\frac{1}{2}(1-e^{-\beta}).
\end{equation}
By a similar derivation, we can obtain
\begin{equation}
\langle L_{4}\rangle=-\frac{1}{2}(1-e^{-\beta}).
\end{equation}

\section{\label{section-histogram-data-Szt-sampled-intervals-Deltat}
The histogram of $S_{z}(t)$ sampled at intervals of $\Delta t$}
In this section, we think about the histogram of $\{S_{z}(n\Delta t)|n=0,1,...,N\}$
for small $\Delta t$ and large $N$.
To simplify the discussion, we assume $\Delta\omega=0$ and $\mbox{\boldmath $S$}(0)=(0,0,0)$.
The time evolution of $\mbox{\boldmath $S$}(t)$ is described as
$\mbox{\boldmath $S$}(t)=(0,0,L_{4}(t))$,
where $L_{4}(t)$ is given by Eq.~(\ref{explicit-form-L4}).

We investigate the time evolution of $L_{4}(t)$ by the following way.
Fixing $\beta$ at a certain value and defining a small time interval $\Delta t$,
we collect $(N+1)$ samples of $L_{4}(n\Delta t)$ for $n=0,1,...,N$,
where $N$ is large enough.
Next, we make a histogram of these $(N+1)$ samples
$\{L_{4}(0),L_{4}(\Delta t),...,L_{4}(N\Delta t)\}$.
We adjust the class interval of bins of the histogram,
so that the line graph of the histogram approaches a smooth curve.

\begin{figure}
\includegraphics[scale=1.2]{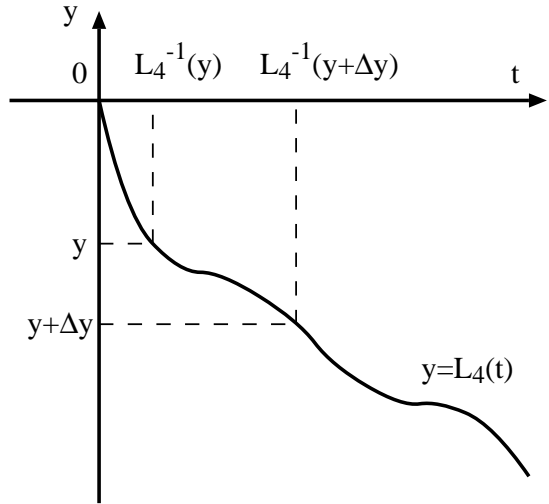}
\caption{The graph of $y=L_{4}(t)$.
The horizontal and vertical lines represent $t$ and $y$ (or $L_{4}$), respectively.
$t$ is in units of $|g|^{-1}$ and $y$ (or $L_{4}$) is a dimensionless quantity.
$t=L_{4}^{-1}(y)$ is an inverse function of $y=L_{4}(t)$.}
\label{Figure04}
\end{figure}

This histogram represents an absolute value of a derivative of the inverse function of $L_{4}(t)$.
As shown in Fig.~\ref{Figure04},
the probability that there is a sample $L_{4}(n\Delta t)$
in the range of the bin of the histogram $[y,y+\Delta y)$ is proportional to
\begin{equation}
\left|
\frac{L_{4}^{-1}(y+\Delta y)-L_{4}^{-1}(y)}{\Delta y}
\right|,
\end{equation}
where $t=L_{4}^{-1}(y)$ is an inverse function of $y=L_{4}(t)$ and $\Delta y$ is a class interval
of the bin.
(Here, we assume $L_{4}^{-1}(y)$ is not a multi-valued function.)
If we take the limit of small $\Delta y$,
this probability reaches
\begin{eqnarray}
&\propto&
\left|
\lim_{\Delta y\to 0}
\frac{L_{4}^{-1}(y+\Delta y)-L_{4}^{-1}(y)}{\Delta y}
\right| \nonumber \\
&=&
\left|
\frac{d}{dy}
L_{4}^{-1}(y)
\right|.
\end{eqnarray}
If $L_{4}^{-1}(y)$ is a multi-valued function,
the probability is proportional to a summation of $|(d/dy)L_{4}^{-1}|$ at points
whose real parts are equal to $y$ on the Riemann surface.

Let us evaluate the histogram of $L_{4}(t)$ in the low temperature limit.
In the low temperature limit,
we can describe $L_{4}(t)$ as
\begin{equation}
y=\lim_{\beta\to\infty}L_{4}(t)
=-\frac{1}{2}(1-\cos 2t).
\end{equation}
Thus, the histogram is proportional to
\begin{eqnarray}
\left|
\frac{dt}{dy}
\right|
&=&
\frac{1}{\sqrt{1-(2y+1)^{2}}} \nonumber \\
&&\quad\quad
\mbox{for $-1\leq y\leq 0$}.
\label{derivative-arcsin}
\end{eqnarray}

Next, we consider the histogram of $L_{4}(t)$ for the high temperature.
If $\beta \ll 1$,
we can expect that $L_{4}(t)$ varies at random around a mean with a certain variance.
Hence, we can approximate the histogram by the probability function of the normal
distribution,
\begin{equation}
\frac{1}{\sqrt{2\pi}\sigma}
e^{-(y-\mu)^{2}/(2\sigma^{2})},
\label{normal-distribution}
\end{equation}
where $\mu$ and $\sigma^{2}$ are a mean and a variance of samples, that is,
\begin{equation}
\mu=\frac{1}{N+1}\sum_{n=0}^{N}L_{4}(n\Delta t),
\end{equation}
and
\begin{equation}
\sigma^{2}=\frac{1}{N+1}\sum_{n=0}^{N}[\mu-L_{4}(n\Delta t)]^{2}.
\end{equation}
If $N\Delta t$ is large enough, $\mu$ is nearly equal to $\langle L_{4}\rangle$.
[From Eq.~(\ref{explicit-form-L4}), we can show $-1\leq L_{4}(t)\leq 0$.
Thus, we can expect that $L_{4}(t)$ varies at random in the range of $[-1,0]$.
However, Eq.~(\ref{normal-distribution}) is defined on $-\infty<y<\infty$.
Here, we neglect this inconsistency.]

\begin{figure}
\includegraphics[scale=1.0]{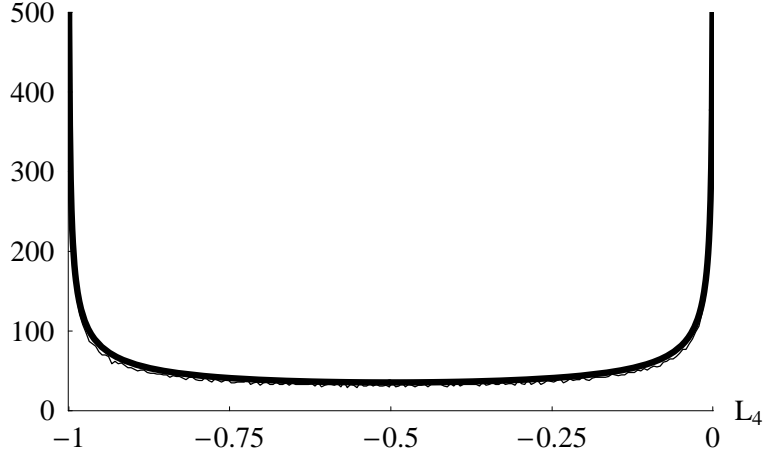}
\caption{The histogram of $\{L_{4}(n\Delta t)|n=0,1,...,9999\}$ where $\beta=10$, $\Delta t=0.05$
and the class interval of each bin is equal to $0.005$.
The horizontal line represents $L_{4}$ that is a dimensionless quantity.
The vertical line represents the number of samples in each bin and it is a dimensionless quantity,
as well.
A thin line graph represents the histogram of samples $\{L_{4}(n\Delta t)\}$.
A thick curve represents the approximate function
$a/\sqrt{1-(2y+1)^{2}}$ where $a=35.5$.
In the figure, the thick curve is lying on the thin line graph and we can hardly distinguish between them.
In the numerical calculation of $L_{4}(t)$ defined in Eq.~(\ref{explicit-form-L4}),
the summation of the index $n$ is carried out up to $1000$.}
\label{Figure05}
\end{figure}

\begin{figure}
\includegraphics[scale=1.0]{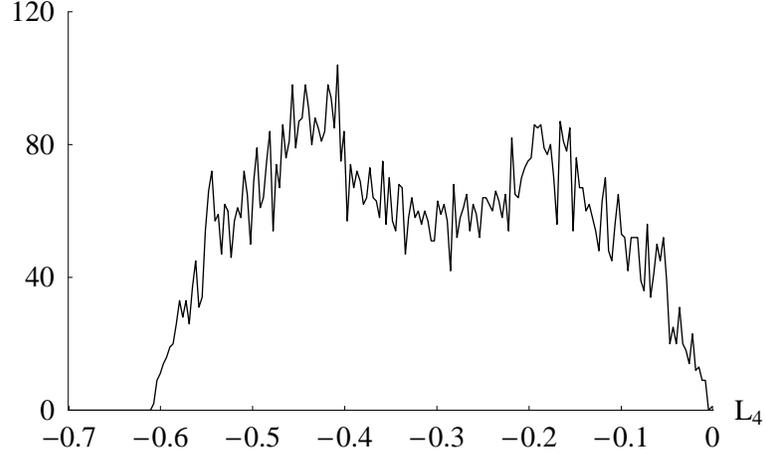}
\caption{The histogram of $\{L_{4}(n\Delta t)|n=0,1,...,9999\}$ where $\beta=1$, $\Delta t=0.05$
and the class interval of each bin is equal to $0.0035$.
The horizontal line represents $L_{4}$ that is a dimensionless quantity.
The vertical line represents the number of samples in each bin and it is a dimensionless quantity,
as well.
In the numerical calculation of $L_{4}(t)$ defined in Eq.~(\ref{explicit-form-L4}),
the summation of the index $n$ is carried out up to $1000$.}
\label{Figure06}
\end{figure}

\begin{figure}
\includegraphics[scale=1.0]{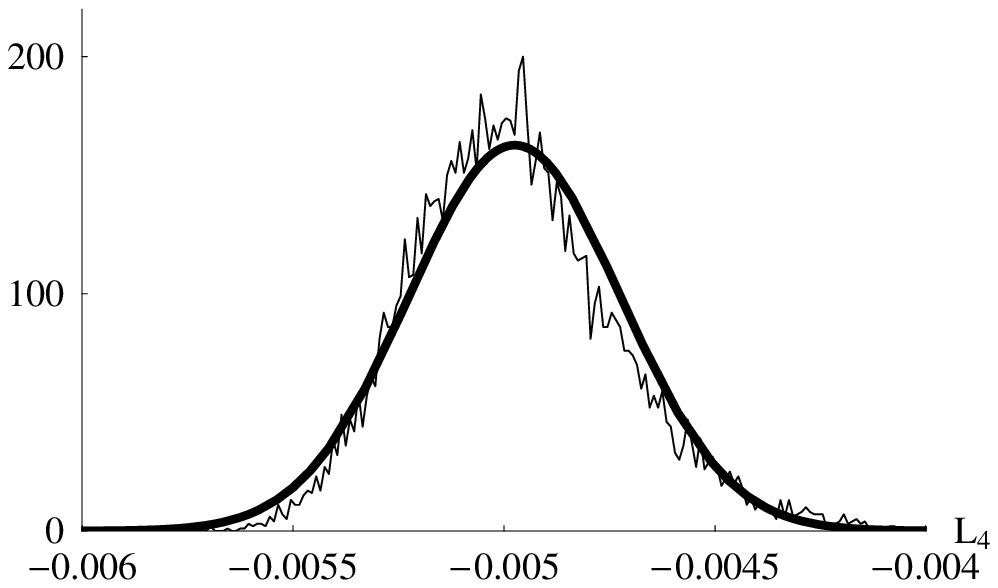}
\caption{The histogram of $\{L_{4}(n\Delta t)|n=0,1,...,9999\}$ where $\beta=0.01$, $\Delta t=0.05$
and the class interval of each bin is equal to $1.0\times 10^{-5}$.
The horizontal line represents $L_{4}$ that is a dimensionless quantity.
The vertical line represents the number of samples in each bin and it is a dimensionless quantity,
as well.
A thin line graph represents the histogram of samples $\{L_{4}(n\Delta t)\}$.
A thick curve represents the approximate function
$[a/(\sqrt{2\pi}\sigma)]\exp[-(y-\mu)^{2}/(2\sigma^{2})]$
where $a=163$, $\mu=-0.00497$ and $\sigma^{2}=6.31\times 10^{-8}$.
(We note $\langle L_{4}\rangle=-0.00498$.)
In the numerical calculation of $L_{4}(t)$ defined in Eq.~(\ref{explicit-form-L4}),
the summation of the index $n$ is carried out up to $1000$.}
\label{Figure07}
\end{figure}

In Figs.~\ref{Figure05}, \ref{Figure06} and \ref{Figure07},
we show the histogram of $L_{4}(t)$ for $\beta=10$, $1$ and $0.01$, respectively.
In Fig.~\ref{Figure05}, Eq.~(\ref{derivative-arcsin}) fits for the histogram well.
In Fig.~\ref{Figure07}, Eq.~(\ref{normal-distribution}) fits for the histogram well, too.
[In Fig.~\ref{Figure07}, the shape of the histogram is not symmetrical.
A slope of the left side is steeper than a slope of the right side.
The author cannot find a physical meaning of this observation.]
The histogram of Fig.~\ref{Figure06} seems to be an intermediate shape of
Figs.~\ref{Figure05} and \ref{Figure07}.

\begin{figure}
\includegraphics[scale=1.0]{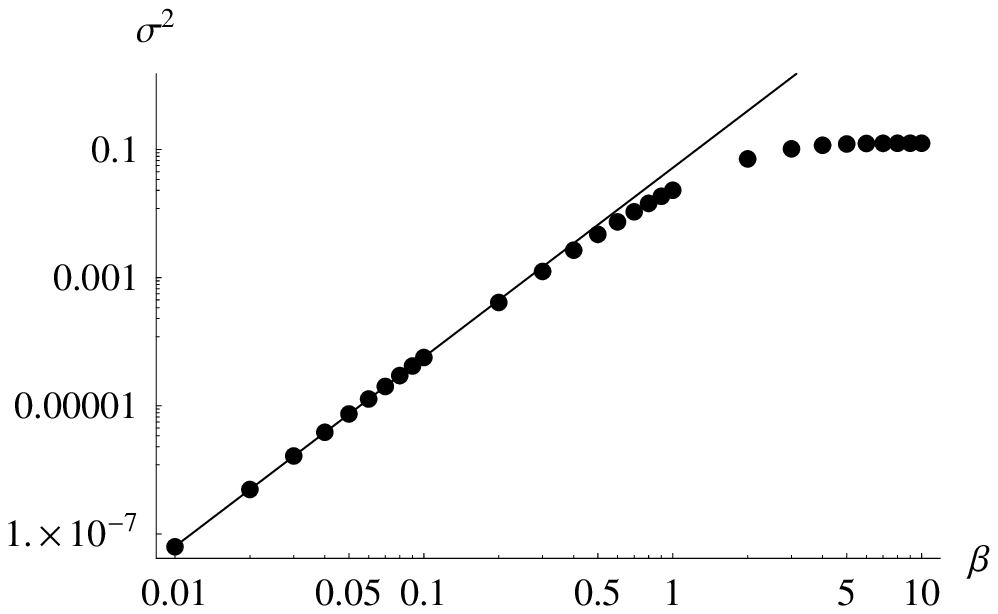}
\caption{Plots of the variance $\sigma^{2}$ of samples $\{L_{4}(n\Delta t)\}$
against $\beta$ for $0.01\leq\beta\leq 10$.
Black circles represent plotted data.
For each black circle, ten thousands samples are taken ($n=0,1,...,9999$)
and we put $\Delta t=0.05$.
The horizontal line represents $\beta$ that is in units of $(\hbar\omega_{0})^{-1}$.
The vertical line represents $\sigma^{2}$ that is a dimensionless quantity.
In both horizontal and vertical lines, ticks are put in the logarithmic scale.
In the range of $0.01\leq\beta\leq 0.1$, plots can be approximated by
$\sigma^{2}(\beta)=c_{1}\beta^{c_{2}}$ where $c_{1}=0.0516$ and $c_{2}=2.95$.
This approximate function is shown as a line graph in the figure.}
\label{Figure08}
\end{figure}

In Fig.~\ref{Figure08}, we plot the variance $\sigma^{2}$ of samples against $\beta$
for $0.01\leq\beta\leq 10$.
For small $\beta$, we can approximate plotted points by
$\sigma^{2}(\beta)=c_{1}\beta^{c_{2}}$ for $0.01\leq\beta\leq 0.1$,
where $c_{1}=0.0516$ and $c_{2}=2.95$.
The author cannot find a reason why the function of $\sigma^{2}(\beta)$ for small $\beta$
has such a simple form.

\section{\label{section-time-average-Bloch-vector-Delta-omega-neq-zero}
The time-average of the Bloch vector for $\Delta\omega\neq 0$}
In this section, we evaluate the time-average of the Bloch vector
for the non-resonant case ($\Delta\omega\neq 0$).
Here, we put $\hbar=1$ ($\beta$ is in units of $\hbar$).
We fix $g$ and $\omega_{0}$ and regard them as constants.
Thus, the model has three variables, $\omega$, $\beta$ and $t$.
From Eqs.~(\ref{A-elements-explicit-form1}), (\ref{equation-time-evolution-Bloch-vector-general}),
(\ref{elements-equation-evolution-Bloch-vector-general}) and (\ref{time-average-Bloch-vector}),
we obtain the following relation:
\begin{equation}
\langle\mbox{\boldmath $S$}\rangle
=
\left(
\begin{array}{ccc}
\langle L_{\Delta\omega}^{(1)}\rangle & \langle L_{\Delta\omega}^{(2)}\rangle & 0 \\
-\langle L_{\Delta\omega}^{(2)}\rangle & \langle L_{\Delta\omega}^{(1)}\rangle & 0 \\
0 & 0 & \langle L_{\Delta\omega}^{(3)}\rangle
\end{array}
\right)
\mbox{\boldmath $S$}(0)
+
\left(
\begin{array}{c}
0 \\
0 \\
\langle L_{\Delta\omega}^{(4)}\rangle
\end{array}
\right),
\label{equation-time-evolution-time-average-Bloch-vector-general}
\end{equation}
where
\begin{eqnarray}
\langle L_{\Delta\omega}^{(1)}\rangle
&=&
(1-e^{-\beta\omega})\lim_{T\to\infty}\frac{1}{T}
\sum_{n=0}^{\infty}
[\frac{n+1}{\sqrt{\tilde{D}(n+1)}}
\sin(\sqrt{\tilde{D}(n+1)}T)
\cos(\sqrt{\tilde{D}(n)}T) \nonumber \\
&&\quad\quad
-
\frac{n}{\sqrt{\tilde{D}(n)}}
\cos(\sqrt{\tilde{D}(n+1)}T)
\sin(\sqrt{\tilde{D}(n)}T)]
e^{-n\beta\omega}, \nonumber \\
\langle L_{\Delta\omega}^{(2)}\rangle
&=&
-\frac{\Delta\omega}{2}
(1-e^{-\beta\omega})\lim_{T\to\infty}\frac{1}{T}
\sum_{n=0}^{\infty}
\frac{\sin(\sqrt{\tilde{D}(n+1)}T)\sin(\sqrt{\tilde{D}(n)}T)}
{\sqrt{\tilde{D}(n+1)\tilde{D}(n)}}
e^{-n\beta\omega}, \nonumber \\
\langle L_{\Delta\omega}^{(3)}\rangle
&=&
(1-e^{-\beta\omega})
\sum_{n=1}^{\infty}
[e^{\beta\omega}\frac{(\Delta\omega/2)^{2}}{\tilde{D}(n)}
+
\frac{1}{2}(e^{\beta\omega}-1)
\frac{g^{2}n}{\tilde{D}(n)}]e^{-n\beta\omega} \nonumber \\
&&\quad\quad
+
\frac{e^{2\beta\omega}-1}{4e^{\beta\omega}}
\lim_{T\to\infty}\frac{1}{T}
\sum_{m=1}^{\infty}
\frac{g^{2}m}{\tilde{D}(m)}
\frac{\sin(2\sqrt{\tilde{D}(m)}T)}{\sqrt{\tilde{D}(m)}}e^{-m\beta\omega}, \nonumber \\
\langle L_{\Delta\omega}^{(4)}\rangle
&=&
(1-e^{-\beta\omega})
\sum_{n=1}^{\infty}
[e^{\beta\omega}\frac{(\Delta\omega/2)^{2}}{\tilde{D}(n)}
+
\frac{1}{2}(e^{\beta\omega}+1)
\frac{g^{2}n}{\tilde{D}(n)}]e^{-n\beta\omega} \nonumber \\
&&\quad\quad
+
\frac{(e^{\beta\omega}-1)^{2}}{4e^{\beta\omega}}
\lim_{T\to\infty}\frac{1}{T}
\sum_{m=1}^{\infty}
\frac{g^{2}m}{\tilde{D}(m)}
\frac{\sin(2\sqrt{\tilde{D}(m)}T)}{\sqrt{\tilde{D}(m)}}e^{-m\beta\omega}
-1.
\label{elements-equation-time-average-evolution-Bloch-vector-general}
\end{eqnarray}

Using the method shown in Sec.~\ref{section-time-average-Bloch-vector-Delta-omega-eq-zero},
we can take the limit $T\to\infty$ in Eq.~(\ref{elements-equation-time-average-evolution-Bloch-vector-general})
and we obtain
\begin{eqnarray}
\langle L_{\Delta\omega}^{(1)}\rangle
&=&
\langle L_{\Delta\omega}^{(2)}\rangle
=0, \nonumber \\
\langle L_{\Delta\omega}^{(3)}\rangle
&=&
(1-e^{-\beta\omega})
\sum_{n=1}^{\infty}
[e^{\beta\omega}\frac{(\Delta\omega/2)^{2}}{\tilde{D}(n)}
+
\frac{1}{2}(e^{\beta\omega}-1)
\frac{g^{2}n}{\tilde{D}(n)}]e^{-n\beta\omega}, \nonumber \\
\langle L_{\Delta\omega}^{(4)}\rangle
&=&
(1-e^{-\beta\omega})
\sum_{n=1}^{\infty}
[e^{\beta\omega}\frac{(\Delta\omega/2)^{2}}{\tilde{D}(n)}
+
\frac{1}{2}(e^{\beta\omega}+1)
\frac{g^{2}n}{\tilde{D}(n)}]e^{-n\beta\omega}-1.
\label{explicit-elements-time-average-evolution-Bloch-vector-general}
\end{eqnarray}

Looking at Eq.~(\ref{explicit-elements-time-average-evolution-Bloch-vector-general}),
we obtain
\begin{equation}
\lim_{\beta\to 0}\langle L_{\Delta\omega}^{(3)}\rangle
=
\lim_{\beta\to 0}\langle L_{\Delta\omega}^{(4)}\rangle
=0,
\end{equation}
\begin{equation}
\lim_{\beta\to\infty}\langle L_{\Delta\omega}^{(3)}\rangle
=
\frac{(\Delta\omega)^{2}+2g^{2}}{(\Delta\omega)^{2}+4g^{2}},
\label{low-temperature-limit-L-Delta-omega-3}
\end{equation}
and
\begin{equation}
\lim_{\beta\to\infty}\langle L_{\Delta\omega}^{(4)}\rangle
=
-\frac{2g^{2}}{(\Delta\omega)^{2}+4g^{2}}.
\label{low-temperature-limit-L-Delta-omega-4}
\end{equation}

The above results imply
$\lim_{\beta\to 0}\langle\mbox{\boldmath $S$}\rangle=\mbox{\boldmath $0$}$
$\forall \mbox{\boldmath $S$}(0)$ and
\begin{eqnarray}
\lim_{\beta\to\infty}\langle\mbox{\boldmath $S$}\rangle
&=&
(0,0,
\frac{[(\Delta\omega)^{2}+2g^{2}]S_{z}(0)-2g^{2}}{(\Delta\omega)^{2}+4g^{2}}) \nonumber \\
&&\quad\quad\forall \mbox{\boldmath $S$}(0).
\label{assymptotic-form-beta-infty-Bloch-vector}
\end{eqnarray}
From Eq.~(\ref{assymptotic-form-beta-infty-Bloch-vector}),
we can say the following:
If we start from $\mbox{\boldmath $S$}(0)=\mbox{\boldmath $0$}$
[the complete mixed state $\rho_{\mbox{\scriptsize A}}(0)=(1/2)\mbox{\boldmath $I$}$]
in the low temperature limit,
we can expect to obtain a slightly purified state of
$\mbox{\boldmath $S$}=(0,0,-2g^{2}/[(\Delta\omega)^{2}+4g^{2}])$
on average after the enough time evolution.

\section{\label{section-entanglement}Entanglement between the atom and the thermal field}
In this section, we consider the evolution of the lower bound of the entanglement of formation
between the atom and the thermal field in the JCM.
In our model defined in Sec.~\ref{section-evolution-Bloch-vector},
the initial state of the atom and the field is separable.
However, because of the Jaynes-Cummings interaction, we can expect
that the entanglement is generated between the atom and the field during the time evolution
and their bipartite state becomes inseparable.
Such a mechanism of entanglement generation is discussed in Refs.~\cite{Bose,Scheel}, as well.
Recently, many researchers have been regarded the JCM as a source of the entanglement
\cite{Kim,Rendell,Larson,El-Orany,Bradler,Yu-Eberly1,Yonac,Yu-Eberly2,Almeida,Sainz}.

Let us pursue the time evolution of the entanglement between the atom and the field
in our model.
To simplify the discussion, we assume $\Delta\omega=0$ and $g>0$.
Furthermore, we assume the atom initially to be in a pure state of
$(1/\sqrt{2})(|0\rangle_{\mbox{\scriptsize A}}+|1\rangle_{\mbox{\scriptsize A}})$,
which implies the Bloch vector $\mbox{\boldmath $S$}(0)=(1,0,0)$ and the density matrix
\begin{equation}
\rho_{\mbox{\scriptsize A}}(0)
=\frac{1}{2}
\left(
\begin{array}{cc}
1 & 1\\
1 & 1
\end{array}
\right). \label{initial-pure-state-100}
\end{equation}
The time evolution of the whole state is described as
\begin{eqnarray}
&&\rho_{\mbox{\scriptsize AF}}(0)
=\rho_{\mbox{\scriptsize A}}(0)\otimes\rho_{\mbox{\scriptsize F}} \nonumber \\
&\longrightarrow&
\rho_{\mbox{\scriptsize AF}}(t)
=U(t)[\rho_{\mbox{\scriptsize A}}(0)\otimes\rho_{\mbox{\scriptsize F}}]U^{\dagger}(t),
\label{time-evolution-atom-field}
\end{eqnarray}
where $\rho_{\mbox{\scriptsize A}}(0)$ is given by Eq.~(\ref{initial-pure-state-100}),
$\rho_{\mbox{\scriptsize F}}$ is given by Eq.~(\ref{field-initial-state}),
and $U(t)$ is given by Eq.~(\ref{explicit-time-evolution-op}).

Here, we are interested in studying the entanglement for the mixed state of the bipartite system
$\mbox{AF}$ (the atom and the field).
[Because $\rho_{\mbox{\scriptsize F}}$ given by Eq.~(\ref{field-initial-state})
is a mixed state, $\rho_{\mbox{\scriptsize AF}}(t)$ in Eq.~(\ref{time-evolution-atom-field}) is also
a mixed state in general.]
Moreover, although the dimension of the system $\mbox{A}$ is finite (the two-dimensional system),
the dimension of the system $\mbox{F}$ is infinite.
Entanglement for such a system is difficult to define.

Some measures of entanglement are proposed at present, for example,
the relative entropy of entanglement, entanglement of formation, and so on.
However, because analytical methods are not found for these measures of the entanglement in general,
it is difficult to compute the value of the entanglement for an arbitrary bipartite state.
However, exceptionally, an explicit formula of the entanglement of formation for an arbitrary $2\times 2$ state
is obtained \cite{Wootters}.
Thus, as the measure of the entanglement, we choose the entanglement of formation.

To investigate the entanglement of the bipartite mixed state $\rho_{\mbox{\scriptsize AF}}(t)$
given by Eq.~(\ref{time-evolution-atom-field}),
we take the following method.
To reduce the dimension of the system $\mbox{F}$ (the field) from an infinite number to a finite number,
we project the entire state of the atom and the field $\rho_{\mbox{\scriptsize AF}}(t)$
onto a subspace whose dimension is given by $2\times 2$ as
\begin{equation}
R_{\mbox{\scriptsize AF}}(t)
=
(|0\rangle_{\mbox{\scriptsize F}}{}_{\mbox{\scriptsize F}}\langle 0|
+
|1\rangle_{\mbox{\scriptsize F}}{}_{\mbox{\scriptsize F}}\langle 1|)
\rho_{\mbox{\scriptsize AF}}(t)
(|0\rangle_{\mbox{\scriptsize F}}{}_{\mbox{\scriptsize F}}\langle 0|
+
|1\rangle_{\mbox{\scriptsize F}}{}_{\mbox{\scriptsize F}}\langle 1|),
\label{definition-density-matrix-RAF}
\end{equation}
where $\{|0\rangle_{\mbox{\scriptsize F}},|1\rangle_{\mbox{\scriptsize F}}\}$
are the photon number states.
(We consider the subspace spanned by the basis vectors
$\{|i\rangle_{\mbox{\scriptsize A}}\otimes|j\rangle_{\mbox{\scriptsize F}}:
i,j\in\{0,1\}\}$.)
Because this operation (the projection) is carried out only in the system $\mbox{F}$ locally,
the entanglement never increases.
Hence, the entanglement of $R_{\mbox{\scriptsize AF}}(t)$ is a lower bound of the entanglement
of $\rho_{\mbox{\scriptsize AF}}(t)$.

After slightly long calculation, we can obtain $R_{\mbox{\scriptsize AF}}(t)$ in the form of
a matrix which is represented in the basis vectors
$\{|0\rangle_{\mbox{\scriptsize A}}|0\rangle_{\mbox{\scriptsize F}},
|0\rangle_{\mbox{\scriptsize A}}|1\rangle_{\mbox{\scriptsize F}},
|1\rangle_{\mbox{\scriptsize A}}|0\rangle_{\mbox{\scriptsize F}},
|1\rangle_{\mbox{\scriptsize A}}|1\rangle_{\mbox{\scriptsize F}}\}$,
\begin{equation}
R_{\mbox{\scriptsize AF}}(t)
=\frac{1}{2}(1-e^{-\beta})
\left(
\begin{array}{cccc}
r_{00,00} & r_{00,01} & r_{00,10} & r_{00,11} \\
r_{00,01}^{\ast} & r_{01,01} & r_{01,10} & r_{01,11} \\
r_{00,10}^{\ast} & r_{01,10}^{\ast} & r_{10,10} & r_{10,11} \\
r_{00,11}^{\ast} & r_{01,11}^{\ast} & r_{10,11}^{\ast} & r_{11,11}
\end{array}
\right),
\label{explicit-form-RAF}
\end{equation}
where
\begin{eqnarray}
r_{00,00}&=&\cos^{2}t+e^{-\beta}\sin^{2}t, \nonumber \\
r_{00,01}&=&i e^{-\beta}\sin t \cos\sqrt{2}t, \nonumber \\
r_{00,10}&=&\cos t, \nonumber \\
r_{00,11}&=&-i\sin t\cos t+i e^{-\beta}\sin t\cos t, \nonumber \\
r_{01,01}&=&e^{-\beta}\cos^{2}\sqrt{2}t+e^{-2\beta}\sin^{2}\sqrt{2}t, \nonumber \\
r_{01,10}&=&0, \nonumber \\
r_{01,11}&=&e^{-\beta}\cos t\cos\sqrt{2}t, \nonumber \\
r_{10,10}&=&1, \nonumber \\
r_{10,11}&=&-i\sin t, \nonumber \\
r_{11,11}&=&\sin^{2}t+e^{-\beta}\cos^{2}t, \label{explicit-components-RAF}
\end{eqnarray}
and we replace $\beta\hbar\omega$ with $\beta$ and $gt$ with $t$.
(Here, we assume $g>0$.)

Here, for convenience, we rewrite $R_{\mbox{\scriptsize AF}}(t)$ as follows:
\begin{equation}
R_{\mbox{\scriptsize AF}}(t)
=p_{\mbox{\scriptsize AF}}(t)\sigma_{\mbox{\scriptsize AF}}(t),
\label{definition-RAF-dash}
\end{equation}
where
\begin{eqnarray}
p_{\mbox{\scriptsize AF}}(t)
&=&\mbox{Tr}[R_{\mbox{\scriptsize AF}}(t)] \nonumber \\
&=&(1/4)(1-e^{-\beta})[4+3e^{-\beta}+e^{-2\beta}+e^{-\beta}(1-e^{-\beta})\cos 2\sqrt{2}t],
\label{probability-RAF}
\end{eqnarray}
and
\begin{equation}
\sigma_{\mbox{\scriptsize AF}}(t)
=
\frac{R_{\mbox{\scriptsize AF}}(t)}{\mbox{Tr}[R_{\mbox{\scriptsize AF}}(t)]}.
\label{definition-sigmaAF}
\end{equation}
[$\sigma_{\mbox{\scriptsize AF}}(t)$ is a normalized density matrix of $R_{\mbox{\scriptsize AF}}(t)$,
that is, $\mbox{Tr}[\sigma_{\mbox{\scriptsize AF}}(t)]=1$.]

The original definition of the entanglement of formation of an arbitrary density matrix
$\rho_{\mbox{\scriptsize AB}}$, which is a bipartite state of systems $\mbox{A}$ and $\mbox{B}$, is as follows.
Let us suppose $\rho_{\mbox{\scriptsize AB}}$ is shared by Alice and Bob.
And suppose that, asymptotically as $n\rightarrow\infty$,
Alice and Bob can prepare $(\rho_{\mbox{\scriptsize AB}})^{\otimes n}$
from $k$ Bell pairs using local operations and classical communication.
The entanglement of formation of $\rho_{\mbox{\scriptsize AB}}$ is given by
\begin{equation}
E(\rho_{\mbox{\scriptsize AB}})
=\lim_{n\to\infty}\frac{k_{\mbox{\scriptsize min}}}{n},
\end{equation}
where $k_{\mbox{\scriptsize min}}$ is the minimum of $k$ for given $n$
\cite{Preskill}.

An explicit formula of the entanglement of formation of a $2\times 2$ dimensional (normalized)
bipartite density matrix $\rho_{\mbox{\scriptsize AB}}$ is given as follows.
First, we compute the concurrence of $\rho_{\mbox{\scriptsize AB}}$ from which we can calculate
the entanglement of formation of $\rho_{\mbox{\scriptsize AB}}$.
We define a matrix $\tilde{\rho}_{\mbox{\scriptsize AB}}$ as
\begin{equation}
\tilde{\rho}_{\mbox{\scriptsize AB}}
=(\sigma_{y}\otimes\sigma_{y})\rho_{\mbox{\scriptsize AB}}^{\ast}(\sigma_{y}\otimes\sigma_{y}),
\end{equation}
where $\rho_{\mbox{\scriptsize AB}}^{\ast}$ is the complex conjugate of $\rho_{\mbox{\scriptsize AB}}$
that is represented in the basis vectors of
$\{|i\rangle_{\mbox{\scriptsize A}}|j\rangle_{\mbox{\scriptsize B}}:i,j\in\{0,1\}\}$.
We write the eigenvalues of $\rho_{\mbox{\scriptsize AB}}\tilde{\rho}_{\mbox{\scriptsize AB}}$
as $\lambda_{1}$, $\lambda_{2}$, $\lambda_{3}$ and $\lambda_{4}$,
where $\lambda_{1}\geq\lambda_{2}\geq\lambda_{3}\geq\lambda_{4}(\geq 0)$.
The concurrence $C(\rho_{\mbox{\scriptsize AB}})$ is given by
\begin{equation}
C(\rho_{\mbox{\scriptsize AB}})
=\mbox{max}\{0,\sqrt{\lambda_{1}}-\sqrt{\lambda_{2}}-\sqrt{\lambda_{3}}-\sqrt{\lambda_{4}}\}.
\end{equation}
The entanglement of formation of $\rho_{\mbox{\scriptsize AB}}$ is given by
${\cal E}(C(\rho_{\mbox{\scriptsize AB}}))$, where
\begin{equation}
{\cal E}(C)
=-\frac{1+\sqrt{1-C^{2}}}{2}\log_{2}\frac{1+\sqrt{1-C^{2}}}{2}
-\frac{1-\sqrt{1-C^{2}}}{2}\log_{2}\frac{1-\sqrt{1-C^{2}}}{2}.
\label{2x2-entanglement-of-formation}
\end{equation}

\begin{figure}
\includegraphics[scale=1.0]{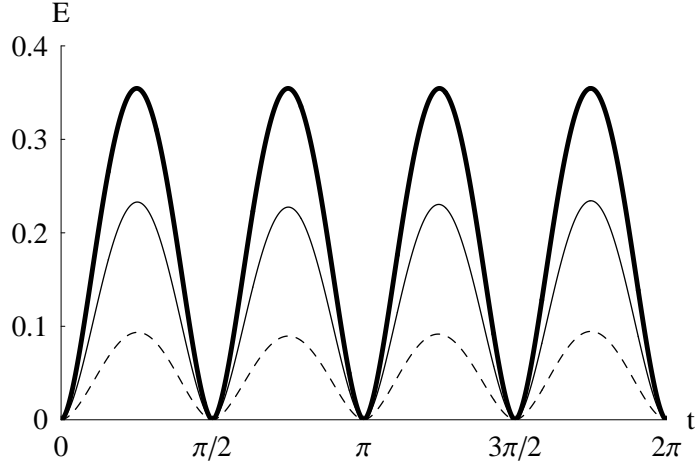}
\caption{The variation of $E(R_{\mbox{\scriptsize AF}}(t))$ against $t\in[0,2\pi]$
with fixed $\beta=10$, $2$ and $1$.
The horizontal and vertical lines represent $t$ (time) and $E$ (entanglement of formation), respectively.
The time $t$ is in units of $g^{-1}$.
The entanglement of formation $E$ is a dimensionless quantity.
A thick curve, a thin curve, and a dashed thin curve represent the variations of
$E(R_{\mbox{\scriptsize AF}}(t))$ for $\beta=10$, $2$, and $1$, respectively.}
\label{Figure09}
\end{figure}

Here, we want to estimate the entanglement of formation of
$R_{\mbox{\scriptsize AF}}(t)=p_{\mbox{\scriptsize AF}}(t)\sigma_{\mbox{\scriptsize AF}}(t)$,
which is not a normalized density matrix.
The entanglement of formation of $R_{\mbox{\scriptsize AF}}(t)$,
$E(R_{\mbox{\scriptsize AF}}(t))$, should be written as
$E(R_{\mbox{\scriptsize AF}}(t))
=p_{\mbox{\scriptsize AF}}(t){\cal E}(C(\sigma_{\mbox{\scriptsize AF}}(t)))$.
Because the analytical form of $\sigma_{\mbox{\scriptsize AF}}(t)$ obtained from
Eqs.~(\ref{explicit-form-RAF}), (\ref{explicit-components-RAF}) and (\ref{definition-sigmaAF})
is very complicated,
it is difficult to obtain $C(\sigma_{\mbox{\scriptsize AF}}(t))$ in an explicit formula.
Thus, we estimate $C(\sigma_{\mbox{\scriptsize AF}}(t))$ 
[and $E(R_{\mbox{\scriptsize AF}}(t))$] numerically.
The variations of $E(R_{\mbox{\scriptsize AF}}(t))$ against $t\in[0,2\pi]$
with fixed $\beta=10$, $2$ and $1$ are shown in Fig.~\ref{Figure09}.

In Fig.~\ref{Figure09}, $E(R_{\mbox{\scriptsize AF}}(t))$ against $t$ seems to vary periodically.
Because we concentrate on the photon number states
$|0\rangle_{\mbox{\scriptsize F}}$ and $|1\rangle_{\mbox{\scriptsize F}}$ only,
the effect of the thermal distribution of the field is neglected.
Looking at Fig.~\ref{Figure09}, we notice that if $\beta$ takes a large value (at a low temperature),
a certain amount of the entanglement arises between the atom and the field.
As $\beta$ becomes smaller (as the temperature becomes higher), the amplitude of oscillation
of $E(R_{\mbox{\scriptsize AF}}(t))$ becomes smaller.
However, this observation does not imply that the entanglement of formation of the entire system
(the atom and the field) becomes smaller as the temperature becomes higher,
because $E(R_{\mbox{\scriptsize AF}}(t))$ is just the lower bound of the entanglement of formation
of $\rho_{\mbox{\scriptsize AF}}(t)$.
From these results, we may regard the thermal JCM as the source of the entanglement.

\section{\label{section-Discussions}Discussions}
In this paper, we examine the dynamics of the Bloch vector of the two-level atom
in the thermal Jaynes-Cummings model (JCM).
In the evolution of the Bloch vector,
for example, if $\Delta\omega=0$,
infinite summation of
$\cos(\sqrt{n+1}t)\cos(\sqrt{n}t)e^{-n\beta}$ and $\cos(2\sqrt{n}t)e^{-n\beta}$
for $n=0,1,2,...$ appears and this makes it difficult
to treat the problem exactly.

In our model, the thermal effects are introduced only in the initial state
of the boson field.
To discuss the thermodynamics of the JCM strictly,
we have to think the grand partition function of the whole system
(the atom and the boson field) and pursue its non-equilibrium time evolution.
Although the JCM has been studied by many researchers,
understanding about the thermal JCM seems not to be enough.

In this paper, we try to obtain a global property that characterizes the confused behavior of the Bloch
vector.
We observe the trajectory of the Bloch vector and take its time-average.
We take the histogram of the $z$ components of the Bloch vector sampled at intervals of $\Delta t$.
However, the author wonders whether these results are good global aspects of the trajectory that is
in complete disorder.

Recently, entanglement generation during the evolution of JCM has been studied from the viewpoint
of the quantum information theory \cite{Bose,Scheel,Kim,Rendell,Larson,El-Orany,Bradler}.
On the other hand, the entanglement sudden death in the JCM is also discussed
\cite{Yu-Eberly1,Yonac,Yu-Eberly2,Almeida,Sainz}.
These matters are explained in Sec.~\ref{section-introduction}.
In Sec.~\ref{section-entanglement}, we consider the evolution of the entanglement of formation between
the atom and the thermal field.
The thermal JCM may become an important source of the entanglement in future.

\end{document}